\documentclass[pre,twocolumn,showpacs,preprintnumbers,amsmath,amssymb,superscriptaddress]{revtex4}

\usepackage{graphicx}
\usepackage{dcolumn}
\usepackage{bm}

\begin{document}

\title{Crystallizing hard-sphere glasses by doping with active particles}

\author{Ran Ni}
\email{rannimail@gmail.com}
\affiliation{%
Van $'$t Hoff Institute for Molecular Sciences, Universiteit van Amsterdam, Science Park 904, 1098 XH Amsterdam, The Netherlands
}%
\affiliation{%
Laboratory of Physical Chemistry and Colloid Science, Wageningen University, Dreijenplein 6, 6703 HB Wageningen, The Netherlands
}%

\author{Martien A. Cohen Stuart}
\affiliation{%
Laboratory of Physical Chemistry and Colloid Science, Wageningen University, Dreijenplein 6, 6703 HB Wageningen, The Netherlands
}%

\author{Marjolein Dijkstra}
\affiliation{Soft Condensed Matter, Utrecht University, Princetonplein 5, 3584 CC Utrecht, The Netherlands}

\author{Peter G. Bolhuis}
\affiliation{%
Van $'$t Hoff Institute for Molecular Sciences, Universiteit van Amsterdam, Science Park 904, 1098 XH Amsterdam, The Netherlands
}%
\begin{abstract}
Crystallization and vitrification are two different routes to form a solid.  Normally these two processes suppress each other, with the glass transition preventing crystallization at high density (or low temperature). This is even true for systems of colloidal hard spheres, which are commonly used as building blocks for novel functional materials with potential applications, e.g. photonic crystals.
 By performing Brownian dynamics simulations of glassy systems
 consisting of mixtures of active and passive hard spheres, we show that the crystallization of such hard-sphere glasses can be dramatically promoted by doping the system with small amounts of active particles. Surprisingly, even hard-sphere glasses of packing fraction up to $\phi = 0.635$ crystallize, which is around $0.5\%$ below the random close packing at $\phi \simeq 0.64$. Our results  suggest 
 a novel way of fabricating crystalline materials from (colloidal) glasses. This is particularly important for materials that get easily kinetically trapped in glassy states, and crystal nucleation hardly occurs.
\end{abstract}

\maketitle
The huge number of important applications associated with crystalline materials have made crystal fabrication a major research theme in the materials science community. The most common route toward a crystal is to supersaturate the corresponding fluid by increasing the density or lowering the temperature, after which the crystal may nucleate. With increasing supersaturation, the driving force for nucleation increases, which lowers the nucleation barrier~\cite{auer2001}. Simultaneously, however, the dynamics of the system also slows down, and at very high supersaturations the metastable fluid phase vitrifies into a glass before crystallization can occur~\cite{zaccarelli2009}. While there are some kinetically arrested glasses that can crystallize slowly via a sequence of stochastic micronucleation events~\cite{sanz2011}, the glass transition generally remains a major obstacle for crystallization of highly supersaturated fluids. 

We discuss here a way to circumvent the kinetic arrest, namely by using active matter. Active matter can be defined as a system of objects capable of continuously converting stored biological or chemical energy into motion. The interest in the dynamics of active matter stems from the wish to understand intriguing self-organization phenomena in nature as featured by bird flocks, bacterial colonies, tissue repair, and the cell cytoskeleton~\cite{revscience2012}. The topic is also growing in chemistry: recent breakthroughs in particle synthesis have enabled the fabrication of artificial colloidal microswimmers that show a high potential for applications in biosensing, drug delivery, etc~\cite{ebbens2010}. 
A number of different active colloidal 
systems have been realized in experiments, such as colloids with magnetic beads that act as artificial flagella~\cite{dreyfus2005}, catalytic Janus particles~\cite{howse2007,erbe2008,palacci2010,baraban2012}, laser-heated metal-capped particles~\cite{volpe2011}, light-activated catalytic colloidal surfers~\cite{palacci2013}, and platinum-loaded stomatocytes \cite{wilson2012}. In contrast to passive colloidal particles that only undergo Brownian motion due to random thermal fluctuations of the solvent, active self-propelled colloids experience an additional force due to internal energy conversion. 
Recent theoretical~\cite{berthier2013} and numerical~\cite{berthier2013prl,ni2013natcom} work has shown that in systems of active hard spheres the glass transition is shifted to densities close to random close packing (RCP), which suggests that the role of activity is to de-vitrify glasses.
In this work, we demonstrate that doping colloidal glasses with small amounts of active particles can significantly enhance the mobility of the passive particles, and speed up the crystallization dynamics. Upon increasing the fraction of active particles, the crystallization pathway switches from spinodal decomposition to nucleation and growth, until too many active particles cause the system to 
adopt a non-equilibrium fluid state. Therefore, there is an optimal fraction of active particles for which the rate of crystallization is maximal. 

We performed event driven Brownian dynamics simulations of a colloidal glass system modeled by $N$  monodisperse hard-sphere particles with a diameter $\sigma$. The hard-sphere system is a simple model for colloidal systems forming glasses~\cite{rmp2010}. Here, we focus on packing fractions above the glass transition, $\phi_g \simeq 0.58$~\cite{zaccarelli2009,rmp2010}.
To prepare the initial  configuration, we use the Lubachevsky-Stillinger algorithm~\cite{LSA} to grow the particles in the simulation box to the packing fraction of interest. 
The total number of particles in the system is fixed at $N=10,000$. We change the fraction of active particles $\alpha$ by randomly selecting $N\alpha$ particles which are made active by applying an self-propelling force $f$ on these particles in the simulations. 
Even though a number of particles in the system are driven and energy is continuously injected to the system, we assume the solvent to be at an equilibrium temperature $T$.
The motion of particle $i$ with position $\mathbf{r}_i$ and orientation $\mathbf{\hat{u}}_i$ can be described via the overdamped Langevin equation given by
\begin{equation}
        \mathbf{\dot{r}}_i(t) = \frac{D_{0}}{k_BT} \left \lbrack -\nabla_i U(t) + \mathbf{\xi}_i(t) + f \mathbf{\hat{u}}_i(t) \right \rbrack,
\end{equation}
where the potential energy $U=\sum_{i<j} U_{HS}(r_{ij})$ is the sum of excluded-volume interactions between all hard spheres with diameter $\sigma$, and $D_{0}$ is the short-time self diffusion coefficient. A stochastic force  with zero mean, $\mathbf{\xi}_i(t)$, describes the  collisions with the solvent molecules, and satisfies $\langle \mathbf{\xi}_i(t)\mathbf{\xi}_j^T(t') \rangle = 2 (k_BT)^2\mathbf{1}\delta_{ij}\delta(t-t')/D_{0}$ with $\mathbf{1}$ the identity matrix. In addition, the self-propulsion of particle $i$ is described by a  constant force $f$ in the direction $\mathbf{\hat{u}}_i(t)$ at time $t$. Note that free swimming speed of the self-propelled particles in dilute suspensions is given by $f D_0/k_B T$. In order to identify the crystalline clusters in the fluid phase, we employ the local bond-order parameter analysis~\cite{bop,nijcp2011,filionjcp2010}.

\begin{figure}[ht!]
\includegraphics[width = 0.45\textwidth]{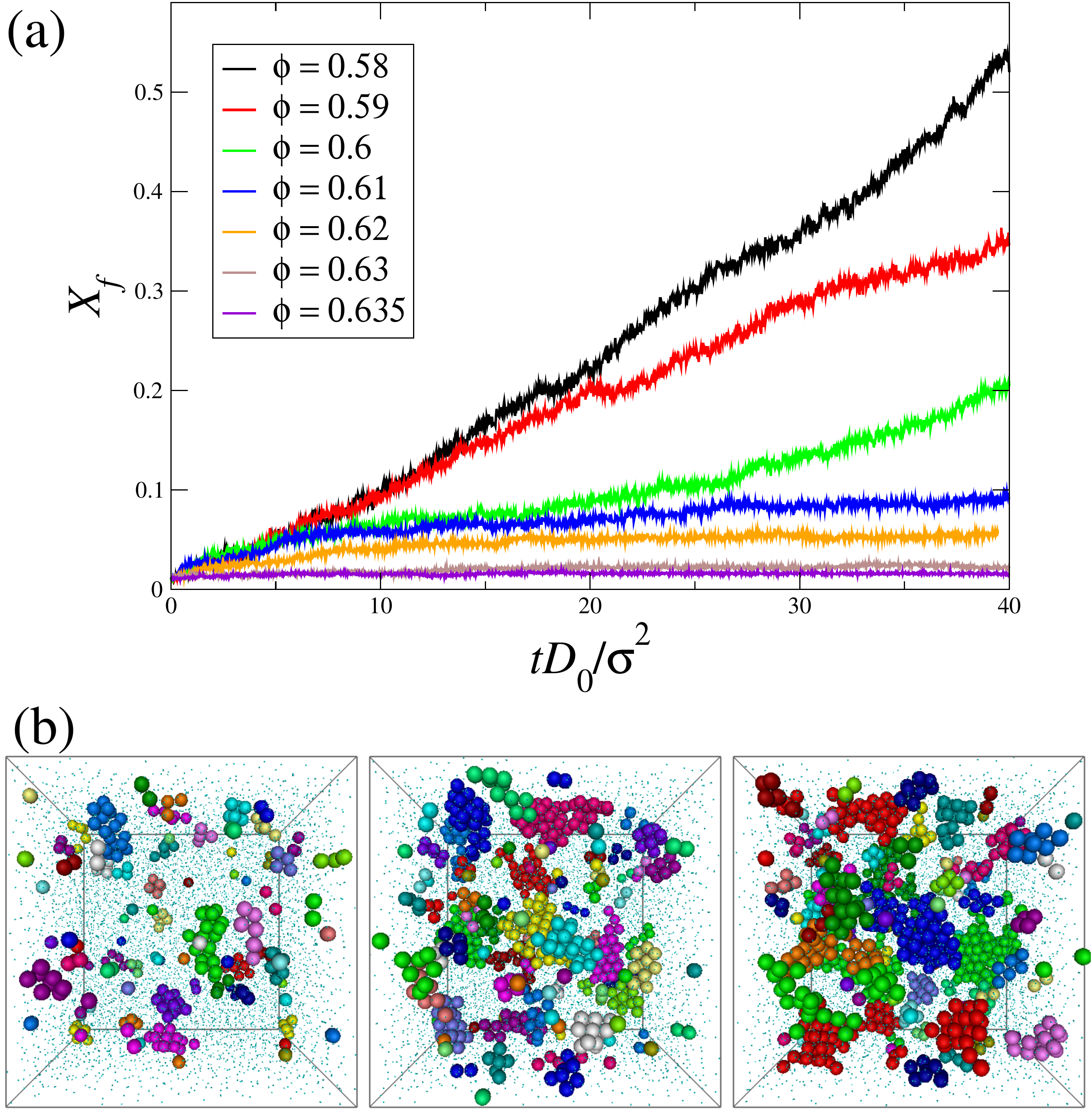}
\caption{\label{fig1} (a) The fraction of crystalline particles $X_{f}$ as a function of time $tD_0/\sigma^2$ in systems of passive Brownian hard-sphere glasses with various packing fractions $0.58 \le \phi \le 0.635$. $D_0$ is the short-time translational self diffusion coefficient. (b) Three snapshots from a typical crystallization trajectory of a passive Brownian hard-sphere glass with $\phi=0.61$ at $tD_0/\sigma^2 = 2,20$ and 40 (from left to right), respectively, where only the particles in crystalline clusters are shown and different colors denote different crystalline clusters. The fluid-like particles are shown as small spheres.}
\end{figure}

Figure~\ref{fig1}a shows the time evolution of the crystalline fraction in hard-sphere glasses for $0.58 \le \phi \le 0.635$, initiated in the disordered state. For packing fractions just above the hard-sphere glass transition $\phi \simeq 0.58$, crystallization to a face-center-cubic (fcc) structure occurs almost immediately via spinodal-decomposition, which is indicated by the immediate increase in crystalline particles without any waiting time. With increasing packing fraction, the crystallization of the hard-sphere glass slows down dramatically due to the emerging glassy dynamics. When $\phi \ge 0.61$, the crystallization is severely suppressed by the slow dynamics. Typical snapshots in Fig.~\ref{fig1}b exemplify the slow growth of crystalline clusters in a dense hard-sphere glass at $\phi=0.61$, in agreement with previous simulation~\cite{sanz2011} and experimental results~\cite{vanmegan1993}.

\begin{figure*}[ht!]
\includegraphics[width = \textwidth]{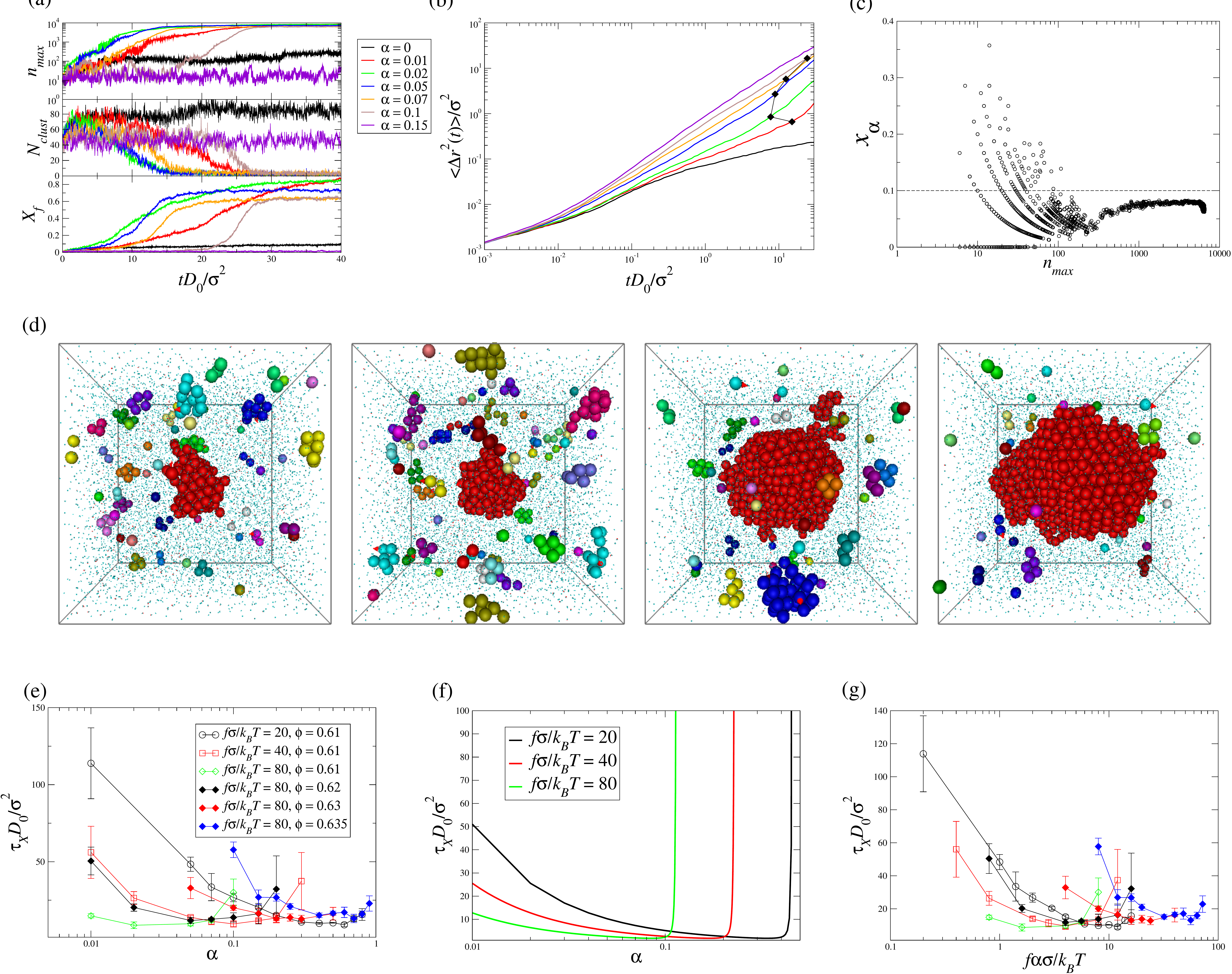}
\caption{\label{fig2} (a) The largest crystalline cluster size $n_{max}$, the number of crystalline clusters $N_{clust}$, and the fraction of crystalline particles $X_{f}$ as a function of time $tD_0/\sigma^2$ in systems of hard-sphere glasses at packing fraction $\phi = 0.61$ doped by various number fractions of active hard spheres $\alpha$ with self-propulsion $f\sigma/k_BT = 80$. (b) The mean square displacement $\langle \Delta r^2 (t)\rangle/\sigma^2$ for the passive particles in the system. The filled diamonds denote the $\langle \Delta r^2 (t)\rangle/\sigma^2$ at $t=\tau_X$. 
(c) Projection of configurations on $n_{max}$ and $x_{\alpha}$ plane for systems of hard-sphere glasses at packing fraction $\phi = 0.61$ doped by $\alpha=0.1$ active hard spheres with self-propulsion $f\sigma/k_BT = 80$. The horizontal dashed line denotes the composition of active particles $\alpha = 0.1$ in the whole system.
(d) Snapshots from a typical trajectory of the nucleation of a single cluster at $\phi=0.61$ with $\alpha = 0.1$ and $f\sigma/k_BT = 80$. From left to right, the time and the corresponding largest cluster size in the system are $(tD_0/\sigma^2,n_{max}) = (17.4,83), (20,168), (22.8,1008)$ and $(24.2,1844)$, respectively. The red arrows on the spheres denote the direction of self-propelling force on the active particles.
(e) Crystallization time $\tau_X D_0/\sigma^2$ as a function of the composition of active particles $\alpha$ in hard-sphere glasses of various packing fractions.
(f) Theoretical prediction of the crystallization time $\tau_X D_0/\sigma^2$ as a function $\alpha$ for different self-propluions $f$ at $\phi = 0.61$. 
(g) Crystallization time $\tau_X D_0/\sigma^2$ as a function $f\alpha \sigma/k_BT$, where the legends of the symbols are the same as in (e).
}
\end{figure*}

Next, we activate a randomly selected small amount of particles in the hard-sphere glass by equipping them with a self-propelling force $f$. This force is applied on the center-of-mass of the particle in the direction $\mathbf{\hat{u}}$. The magnitude of the self-propelling force $f$ is constant, while the orientation $\mathbf{\hat{u}}$ undergoes a free Brownian rotation with a rotational diffusion coefficient $D_r = 3 D_0/\sigma^2$ according to the Stokes-Einstein relationship. Figure~\ref{fig2}a shows the time evolution of the crystalline fraction in a hard-sphere glass of packing fraction $\phi = 0.61$ containing various number fractions of active hard spheres, i.e. $0 \le \alpha \le 0.15$, with $f\sigma/k_BT = 80$. Clearly, by doping the glass with only $1\%$ of active hard spheres, the crystallization dynamics speeds up significantly. Similar to the passive hard-sphere glass, crystallization of the hard-sphere glass with $\alpha = 0.01$ occurs immediately without any waiting time via spinodal decomposition as shown in Fig.~\ref{fig2}a, which can be seen from the coupling between the increase of the largest cluster $n_{max}$ and the decrease of the number of clusters $N_{clust}$. In addition, as shown in Fig.~\ref{fig2}b, 
the mean square displacement of passive hard spheres increases faster in systems with more active particles, which is in agreement with Ref.~\cite{ni2013natcom}.
This suggests that the existence of active particles in the glass enhances the mobility of the passive particles, which assists the coalescence of the clusters and speeds up the crystallization dynamics. 
We define a ``crystallization time'' $\tau_X$ as the time to reach a fraction of crystalline particles $X_f = 0.2$ as in Ref.~\cite{chantal2011}.
Figure~\ref{fig2}e shows that with increasing fraction of active particles $\alpha$ the crystallization time $\tau_X$ decreases to a minimum at $\alpha = 0.02$, after which $\tau_X$ increases again with $\alpha$. 
This  increase 
is due to the fact that the presence of active particles lowers the stability of the resulting crystal phase, thus decreasing the driving force of the phase transition~\cite{lowen2012}. Therefore, the nucleation rate, i.e. the number of critical nuclei per unit volume and time, decreases. 
Indeed, at $\alpha = 0.1$, as shown in Fig.~\ref{fig2}a, the hard-sphere glass first melts into a non-equilibrium fluid phase, which is indicated by the large diffusivity of the particles without any crystalline clusters (Fig. \ref{fig2}a,b). Then the system stays in the non-equilibrium fluid state for a time $t^*D_0/\sigma^2 \simeq 20$, followed by nucleation and growth of a single critical cluster (see a typical sequence of snapshots in Fig.~\ref{fig2}d).
It is surprising that the nucleation and growth of a single nucleus can occur in such a dense hard-sphere glass.
This finding allows us to further study the role of active particles in the crystal nucleation. In Fig.~\ref{fig2}c, we map the configurations onto the $n_{max}-x_{\alpha}$ plane, in which $n_{max}$ and $x_{\alpha}$ are the size of the largest cluster and the fraction of active particles in this cluster, respectively.
For small clusters, the fraction of active particles fluctuates strongly between 0 and 0.4, but when it increases to around the critical size, i.e. $100 < n_{max}^* < 200$, the fraction of active particles in the nuclei remains below $\alpha$,
suggesting that the critical nucleus comprises mainly of passive hard spheres. Therefore, the active particles in the system mainly act as ``stirrers''  or ``mixers'',  speeding up the mobility of the passive particles without actually initiating the crystal nucleation.

After further increasing the active particle fraction to $\alpha > 0.1$ the system stays in the non-equilibrium fluid phase, and nucleation remains a rare event.  In fact, it may even be hampered completely, since the large amount of active particles can also melt the crystal into a fluid phase~\cite{lowen2012}.
Hence, the enhanced  crystallization of hard-sphere glasses by doping with active particles is the result of a competition between mobilizing (stirring)  the glass and destabilizing the resulting crystals, yielding a maximal crystallization speed at around $\alpha^* \simeq 0.02$ with $f\sigma/k_BT = 80$ at $\phi = 0.61$.

The influence of self-propulsion on the crystallization time  is summarized in Fig.~\ref{fig2}e. 
Increasing the fraction of active particles in hard-sphere glasses from $\alpha = 0$ always decreases the crystallization time $\tau_X$, reaching a minimum at some optimal doping fraction, after which $\tau_X$ increases again with $\alpha$. 
The open symbols in Fig.~\ref{fig2}e show that  the optimal composition of active particles decreases with increasing magnitude $f$ of the self-propulsion force of active particles, i.e.
to induce coalescence of the crystalline clusters in the glass requires a decreasing amount of stronger active particles.
Similarly, as shown by the solid symbols in Fig.~\ref{fig2}e, for $f\sigma/k_BT = 80$ the optimal composition of active particles increases with packing fraction because a larger amount of stirring by active particles is required to crystallize denser glasses.
Strikingly, by doping  with only $10\%$ active particles with $f\sigma/k_BT = 80$, we have succeeded in crystallizing a hard-sphere glass at $\phi=0.635$, a packing fraction for which crystallization has never been observed before~\cite{zaccarelli2009,sanz2011}.  The measured crystallization time is also shown in Fig.~\ref{fig2}e.

We can rationalize the above behaviour by adapting  classical nucleation theory (CNT) to our system. While CNT is clearly derived for quasi-equilibrium systems, we assume that nucleation in a non-equilibrium system is still governed by the same fundamental physics. 
 The two main factors in CNT, the nucleation barrier and the kinetic prefactor,  will be influenced by the presence of active particles. The kinetic prefactor will increase, because of the active particles, and it is likely that the nucleation barrier will increase, as the driving force for crystallization will be lower due to the active particles. 
A more quantitative analysis can be put forward as follows. According to CNT, the crystallization rate as a function of  $\alpha$ and $f$ has the form  
\begin{equation}
k(\alpha,f) = D(\alpha,f) e^{-\Delta G(\alpha,f)/ k_BT} \simeq 1/\tau_X,
\end{equation}
 with $D(\alpha,f)$ a kinetic prefactor proportional to the diffusion and 
\begin{equation}
\Delta G(\alpha,f) = \frac{16 \pi \gamma^3}{3\rho^2 \Delta \mu(\alpha,f)^2}
\end{equation} 
is the nucleation barrier height, with $\gamma$ the surface tension, $\rho$ the density, and $\Delta \mu(\alpha,f)=\mu_{sol} - \mu_{liq}$ the driving force~\cite{nithesis}. 
The dependence on $\alpha$ and $f$ is probably very complicated, but as a first approximation we can take a simple linear function~\cite{Jepson2013}, $D(\alpha,f) = D^* + c_1 f \alpha $ and $\Delta \mu(\alpha,f) = \Delta \mu_0 - c_2 f \alpha $, with $D^*$ the passive kinetic prefactor, and $\Delta \mu_0$ the passive driving force for crystallization.
For our system $\phi=0.61$, the diffusivity $D^* \simeq 0$, $\gamma=0.7k_BT/\sigma^2$ and $\Delta \mu_0 = 14.58 k_B T$ (from  the Carnahan Starling and Hall equation of state~\cite{halleos}). Setting parameters $c_1 =0.1$ and $c_2=1.5$ the predicted crystallization rates are plotted in Fig.~\ref{fig2}f as a function of $\alpha$, for different $f\sigma/k_B T=20,40,80$. Clearly the qualitative behaviour of our simulations are reproduced: we find first an increase in crystallization rate ($1/\tau_X$) due to the enhanced diffusion, followed by a decrease due to a higher nucleation barrier caused by a reduced driving force.  
However, as shown in Fig.~\ref{fig2}g, the crystallization time for different $f$ cannot simply be scaled onto a single curve by plotting $\tau_X$ as a function of $f \alpha$, and with larger $f$, the optimal $f \alpha$ for the fastest crystallization decreases. This suggests that the effect of active particle on the crystallization can not be described via the simple linear combination $f \alpha$, and  further investigation is required.

In conclusion, by performing event-driven Brownian dynamics simulations, we systematically study the crystallization of hard-sphere glasses consisting of a mixture of passive and active hard spheres. 
Our results can be summarized as follows: 
1) doping hard-sphere glasses with a small amount of active particles enhances the mobility of the passive particles, which assists the coalescence of the crystalline clusters and speeds up the crystallization dynamics via spinodal decomposition; 2) upon increasing the fraction of active particles further, the crystallization speed reaches a maximum, beyond which the glass first melts into a non-equilibrium fluid, and the crystal may form via nucleation and growth afterwards.
These results can be reasonably well explain by a modified CNT.
By doping with active particles,
we are able to  crystallize hard-sphere glasses up to $\phi = 0.635$, around $0.5\%$ below the RCP limit $\sim 0.64$, for which no crystallization has ever been observed before. 
When the density of a colloidal glass approaches the RCP, the pressure of the system diverges, and the driving force of the crystallization increases to infinity.   
Since we find that the fraction of active particles in the critical nuclei is lower than that in the bulk phase, doping with a tiny amount of strong active particles enhances the mobility of passive particles without changing significantly the stability of the crystal phase, which enables the crystallization of glasses at densities close to the RCP. 
Hence, we expect that this crystallization method can be employed for most colloidal glasses as long as the thermodynamically stable phase is a crystal. For instance, in systems of binary hard spheres, the crystallization has become highly challenging, and we hope our method can be future employed to help the crystallization of binary hard-sphere crystals. Moreover, we wish to note that if the active particles are much smaller  than the passive particles, the situation can be much more complicated, since it has been found that the small active particles can produce giant long range effect interactions between the large passive particles in the system~\cite{ni2014}.
Although most experimental studies correspond to low activities $f\sigma/k_B T < 10$~\cite{howse2007,palacci2010,volpe2011,schwarz-linek2012,wilson2012}, light-activated colloids \cite{palacci2013}, the catalytic Janus particles \cite{baraban2012}, and the particles with a artificial magnetic flagella \cite{dreyfus2005} are capable of producing self-propulsions as high as $f \sigma/k_B T \simeq 20$, 50, 80, respectively, making our findings highly relevant for these systems. Moreover, one can also use optical tweezers to actively move small amounts of colloidal particles in the glass with large forces~\cite{sprakel2013}, which should have a profound effect on the crystallization dynamics of colloidal glasses. In this work, we neglected the effect of hydrodynamics, which could be an important direction for future investigation. However, it has been found that Brownian dynamics simulations without explicit hydrodynamics can reproduce most of structures and patterns observed in experiments~\cite{palacci2013,lowen2012}, which suggests that the method used in this work can capture the essential physics in the dynamic assembly of active colloidal swimmers.

Our results suggest a new way of fabricating crystalline materials from glasses, and it may be particularly important for the crystallization of photonic crystals~\cite{nanooptics,ab13}.

\begin{acknowledgments} R.N. and M.A.C.S acknowledge financial support from the European Research Council through Advanced Grant 267254 (BioMate).
This work is part of the research programme VICI 700.58.442, which is financed by the Netherlands Organization for Scientific Research (NWO).
\end{acknowledgments}


\begin{thebibliography}{32}
\expandafter\ifx\csname natexlab\endcsname\relax\def\natexlab#1{#1}\fi
\expandafter\ifx\csname bibnamefont\endcsname\relax
  \def\bibnamefont#1{#1}\fi
\expandafter\ifx\csname bibfnamefont\endcsname\relax
  \def\bibfnamefont#1{#1}\fi
\expandafter\ifx\csname citenamefont\endcsname\relax
  \def\citenamefont#1{#1}\fi
\expandafter\ifx\csname url\endcsname\relax
  \def\url#1{\texttt{#1}}\fi
\expandafter\ifx\csname urlprefix\endcsname\relax\def\urlprefix{URL }\fi
\providecommand{\bibinfo}[2]{#2}
\providecommand{\eprint}[2][]{\url{#2}}

\bibitem[{\citenamefont{Auer and Frenkel}(2001)}]{auer2001}
\bibinfo{author}{\bibfnamefont{S.}~\bibnamefont{Auer}} \bibnamefont{and}
  \bibinfo{author}{\bibfnamefont{D.}~\bibnamefont{Frenkel}},
  \bibinfo{journal}{Nature (London)} \textbf{\bibinfo{volume}{409}},
  \bibinfo{pages}{1020} (\bibinfo{year}{2001}).

\bibitem[{\citenamefont{Zaccarelli et~al.}(2009)\citenamefont{Zaccarelli,
  Valeriani, Sanz, Poon, Cates, and Pusey}}]{zaccarelli2009}
\bibinfo{author}{\bibfnamefont{E.}~\bibnamefont{Zaccarelli}},
  \bibinfo{author}{\bibfnamefont{C.}~\bibnamefont{Valeriani}},
  \bibinfo{author}{\bibfnamefont{E.}~\bibnamefont{Sanz}},
  \bibinfo{author}{\bibfnamefont{W.~C.~K.} \bibnamefont{Poon}},
  \bibinfo{author}{\bibfnamefont{M.~E.} \bibnamefont{Cates}}, \bibnamefont{and}
  \bibinfo{author}{\bibfnamefont{P.~N.} \bibnamefont{Pusey}},
  \bibinfo{journal}{Phys. Rev. Lett.} \textbf{\bibinfo{volume}{103}},
  \bibinfo{pages}{135704} (\bibinfo{year}{2009}).

\bibitem[{\citenamefont{Sanz et~al.}(2011)\citenamefont{Sanz, Valeriani,
  Zaccarelli, Poon, Pusey, and Cates}}]{sanz2011}
\bibinfo{author}{\bibfnamefont{E.}~\bibnamefont{Sanz}},
  \bibinfo{author}{\bibfnamefont{C.}~\bibnamefont{Valeriani}},
  \bibinfo{author}{\bibfnamefont{E.}~\bibnamefont{Zaccarelli}},
  \bibinfo{author}{\bibfnamefont{W.~C.~K.} \bibnamefont{Poon}},
  \bibinfo{author}{\bibfnamefont{P.~N.} \bibnamefont{Pusey}}, \bibnamefont{and}
  \bibinfo{author}{\bibfnamefont{M.~E.} \bibnamefont{Cates}},
  \bibinfo{journal}{Phys. Rev. Lett.} \textbf{\bibinfo{volume}{106}},
  \bibinfo{pages}{215701} (\bibinfo{year}{2011}).

\bibitem[{\citenamefont{Gonzalez-Rodriguez
  et~al.}(2012)\citenamefont{Gonzalez-Rodriguez, Guevorkian, Douezan, and
  Brochard-Wyart}}]{revscience2012}
\bibinfo{author}{\bibfnamefont{D.}~\bibnamefont{Gonzalez-Rodriguez}},
  \bibinfo{author}{\bibfnamefont{K.}~\bibnamefont{Guevorkian}},
  \bibinfo{author}{\bibfnamefont{S.}~\bibnamefont{Douezan}}, \bibnamefont{and}
  \bibinfo{author}{\bibfnamefont{F.}~\bibnamefont{Brochard-Wyart}},
  \bibinfo{journal}{Science} \textbf{\bibinfo{volume}{338}},
  \bibinfo{pages}{910} (\bibinfo{year}{2012}).

\bibitem[{\citenamefont{Ebbens and Howse}(2010)}]{ebbens2010}
\bibinfo{author}{\bibfnamefont{S.}~\bibnamefont{Ebbens}} \bibnamefont{and}
  \bibinfo{author}{\bibfnamefont{J.}~\bibnamefont{Howse}},
  \bibinfo{journal}{Soft Matter} \textbf{\bibinfo{volume}{6}},
  \bibinfo{pages}{726} (\bibinfo{year}{2010}).

\bibitem[{\citenamefont{Dreyfus et~al.}(2005)\citenamefont{Dreyfus, Baudry,
  Roper, Fermigier, Stone, and Bibette}}]{dreyfus2005}
\bibinfo{author}{\bibfnamefont{R.}~\bibnamefont{Dreyfus}},
  \bibinfo{author}{\bibfnamefont{J.}~\bibnamefont{Baudry}},
  \bibinfo{author}{\bibfnamefont{M.}~\bibnamefont{Roper}},
  \bibinfo{author}{\bibfnamefont{M.}~\bibnamefont{Fermigier}},
  \bibinfo{author}{\bibfnamefont{H.}~\bibnamefont{Stone}}, \bibnamefont{and}
  \bibinfo{author}{\bibfnamefont{J.}~\bibnamefont{Bibette}},
  \bibinfo{journal}{Nature} \textbf{\bibinfo{volume}{437}},
  \bibinfo{pages}{862} (\bibinfo{year}{2005}).

\bibitem[{\citenamefont{Howse et~al.}(2007)\citenamefont{Howse, Jones, Ryan,
  Gough, Vafabakhsh, and Golestanian}}]{howse2007}
\bibinfo{author}{\bibfnamefont{J.~R.} \bibnamefont{Howse}},
  \bibinfo{author}{\bibfnamefont{R.~A.~L.} \bibnamefont{Jones}},
  \bibinfo{author}{\bibfnamefont{A.~J.} \bibnamefont{Ryan}},
  \bibinfo{author}{\bibfnamefont{T.}~\bibnamefont{Gough}},
  \bibinfo{author}{\bibfnamefont{R.}~\bibnamefont{Vafabakhsh}},
  \bibnamefont{and}
  \bibinfo{author}{\bibfnamefont{R.}~\bibnamefont{Golestanian}},
  \bibinfo{journal}{Phys. Rev. Lett.} \textbf{\bibinfo{volume}{99}},
  \bibinfo{pages}{048102} (\bibinfo{year}{2007}).

\bibitem[{\citenamefont{Erbe et~al.}(2008)\citenamefont{Erbe, Zientara,
  Baraban, Kreidler, and Leiderer}}]{erbe2008}
\bibinfo{author}{\bibfnamefont{A.}~\bibnamefont{Erbe}},
  \bibinfo{author}{\bibfnamefont{M.}~\bibnamefont{Zientara}},
  \bibinfo{author}{\bibfnamefont{L.}~\bibnamefont{Baraban}},
  \bibinfo{author}{\bibfnamefont{C.}~\bibnamefont{Kreidler}}, \bibnamefont{and}
  \bibinfo{author}{\bibfnamefont{P.}~\bibnamefont{Leiderer}},
  \bibinfo{journal}{J. Phys.: Condens. Matter} \textbf{\bibinfo{volume}{20}},
  \bibinfo{pages}{404215} (\bibinfo{year}{2008}).

\bibitem[{\citenamefont{Palacci et~al.}(2010)\citenamefont{Palacci,
  Cottin-Bizonne, Ybert, and Bocquet}}]{palacci2010}
\bibinfo{author}{\bibfnamefont{J.}~\bibnamefont{Palacci}},
  \bibinfo{author}{\bibfnamefont{C.}~\bibnamefont{Cottin-Bizonne}},
  \bibinfo{author}{\bibfnamefont{C.}~\bibnamefont{Ybert}}, \bibnamefont{and}
  \bibinfo{author}{\bibfnamefont{L.}~\bibnamefont{Bocquet}},
  \bibinfo{journal}{Phys. Rev. Lett.} \textbf{\bibinfo{volume}{105}},
  \bibinfo{pages}{088304} (\bibinfo{year}{2010}).

\bibitem[{\citenamefont{Baraban et~al.}(2012)\citenamefont{Baraban,
  Tasinkevych, Popescu, Sanchez, Dietrich, and Schmidt}}]{baraban2012}
\bibinfo{author}{\bibfnamefont{L.}~\bibnamefont{Baraban}},
  \bibinfo{author}{\bibfnamefont{M.}~\bibnamefont{Tasinkevych}},
  \bibinfo{author}{\bibfnamefont{M.}~\bibnamefont{Popescu}},
  \bibinfo{author}{\bibfnamefont{S.}~\bibnamefont{Sanchez}},
  \bibinfo{author}{\bibfnamefont{S.}~\bibnamefont{Dietrich}}, \bibnamefont{and}
  \bibinfo{author}{\bibfnamefont{O.}~\bibnamefont{Schmidt}},
  \bibinfo{journal}{Soft Matter} \textbf{\bibinfo{volume}{8}},
  \bibinfo{pages}{48} (\bibinfo{year}{2012}).

\bibitem[{\citenamefont{Volpe et~al.}(2011)\citenamefont{Volpe, Buttinoni,
  Vogt, K\"ummerer, and Bechinger}}]{volpe2011}
\bibinfo{author}{\bibfnamefont{G.}~\bibnamefont{Volpe}},
  \bibinfo{author}{\bibfnamefont{I.}~\bibnamefont{Buttinoni}},
  \bibinfo{author}{\bibfnamefont{D.}~\bibnamefont{Vogt}},
  \bibinfo{author}{\bibfnamefont{H.}~\bibnamefont{K\"ummerer}},
  \bibnamefont{and}
  \bibinfo{author}{\bibfnamefont{C.}~\bibnamefont{Bechinger}},
  \bibinfo{journal}{Soft Matter} \textbf{\bibinfo{volume}{7}},
  \bibinfo{pages}{8810} (\bibinfo{year}{2011}).

\bibitem[{\citenamefont{Palacci et~al.}(2013)\citenamefont{Palacci, Sacanna,
  Steinberg, Pine, and Chaikin}}]{palacci2013}
\bibinfo{author}{\bibfnamefont{J.}~\bibnamefont{Palacci}},
  \bibinfo{author}{\bibfnamefont{S.}~\bibnamefont{Sacanna}},
  \bibinfo{author}{\bibfnamefont{A.~P.} \bibnamefont{Steinberg}},
  \bibinfo{author}{\bibfnamefont{D.~J.} \bibnamefont{Pine}}, \bibnamefont{and}
  \bibinfo{author}{\bibfnamefont{P.~M.} \bibnamefont{Chaikin}},
  \bibinfo{journal}{Science} \textbf{\bibinfo{volume}{339}},
  \bibinfo{pages}{936} (\bibinfo{year}{2013}).

\bibitem[{\citenamefont{Wilson et~al.}(2012)\citenamefont{Wilson, Nolte, and
  van Hest}}]{wilson2012}
\bibinfo{author}{\bibfnamefont{D.~A.} \bibnamefont{Wilson}},
  \bibinfo{author}{\bibfnamefont{R.~J.~M.} \bibnamefont{Nolte}},
  \bibnamefont{and} \bibinfo{author}{\bibfnamefont{J.~C.~M.} \bibnamefont{van
  Hest}}, \bibinfo{journal}{Nature Chemistry} \textbf{\bibinfo{volume}{4}},
  \bibinfo{pages}{268} (\bibinfo{year}{2012}).

\bibitem[{\citenamefont{Berthier and Kurchan}(2013)}]{berthier2013}
\bibinfo{author}{\bibfnamefont{L.}~\bibnamefont{Berthier}} \bibnamefont{and}
  \bibinfo{author}{\bibfnamefont{J.}~\bibnamefont{Kurchan}},
  \bibinfo{journal}{Nature Phys.} \textbf{\bibinfo{volume}{9}},
  \bibinfo{pages}{310} (\bibinfo{year}{2013}).

\bibitem[{\citenamefont{Berthier}(2013)}]{berthier2013prl}
\bibinfo{author}{\bibfnamefont{L.}~\bibnamefont{Berthier}},
  \bibinfo{journal}{arXiv:1307.0704}  (\bibinfo{year}{2013}).

\bibitem[{\citenamefont{Ni et~al.}(2013)\citenamefont{Ni, Cohen-Stuart, and
  Dijkstra}}]{ni2013natcom}
\bibinfo{author}{\bibfnamefont{R.}~\bibnamefont{Ni}},
  \bibinfo{author}{\bibfnamefont{M.}~\bibnamefont{Cohen-Stuart}},
  \bibnamefont{and} \bibinfo{author}{\bibfnamefont{M.}~\bibnamefont{Dijkstra}},
  \bibinfo{journal}{Nature Communications} \textbf{\bibinfo{volume}{4}},
  \bibinfo{pages}{2704} (\bibinfo{year}{2013}).

\bibitem[{\citenamefont{Parisi and Zamponi}(2010)}]{rmp2010}
\bibinfo{author}{\bibfnamefont{G.}~\bibnamefont{Parisi}} \bibnamefont{and}
  \bibinfo{author}{\bibfnamefont{F.}~\bibnamefont{Zamponi}},
  \bibinfo{journal}{Rev. Mod. Phys.} \textbf{\bibinfo{volume}{82}},
  \bibinfo{pages}{789} (\bibinfo{year}{2010}).

\bibitem[{\citenamefont{Lubachevsky and Stillinger}(1990)}]{LSA}
\bibinfo{author}{\bibfnamefont{B.~D.} \bibnamefont{Lubachevsky}}
  \bibnamefont{and} \bibinfo{author}{\bibfnamefont{F.~H.}
  \bibnamefont{Stillinger}}, \bibinfo{journal}{J. Stat. Phys.}
  \textbf{\bibinfo{volume}{60}}, \bibinfo{pages}{561} (\bibinfo{year}{1990}).

\bibitem[{\citenamefont{Steinhardt et~al.}(1983)\citenamefont{Steinhardt,
  Nelson, and Ronchetti}}]{bop}
\bibinfo{author}{\bibfnamefont{P.~J.} \bibnamefont{Steinhardt}},
  \bibinfo{author}{\bibfnamefont{D.~R.} \bibnamefont{Nelson}},
  \bibnamefont{and}
  \bibinfo{author}{\bibfnamefont{M.}~\bibnamefont{Ronchetti}},
  \bibinfo{journal}{Phys. Rev. B} \textbf{\bibinfo{volume}{28}},
  \bibinfo{pages}{784} (\bibinfo{year}{1983}).

\bibitem[{\citenamefont{Ni and Dijkstra}(2011)}]{nijcp2011}
\bibinfo{author}{\bibfnamefont{R.}~\bibnamefont{Ni}} \bibnamefont{and}
  \bibinfo{author}{\bibfnamefont{M.}~\bibnamefont{Dijkstra}},
  \bibinfo{journal}{J. Chem. Phys.} \textbf{\bibinfo{volume}{134}},
  \bibinfo{pages}{034501} (\bibinfo{year}{2011}).

\bibitem[{\citenamefont{Filion et~al.}(2010)\citenamefont{Filion, Hermes, Ni,
  and Dijkstra}}]{filionjcp2010}
\bibinfo{author}{\bibfnamefont{L.}~\bibnamefont{Filion}},
  \bibinfo{author}{\bibfnamefont{M.}~\bibnamefont{Hermes}},
  \bibinfo{author}{\bibfnamefont{R.}~\bibnamefont{Ni}}, \bibnamefont{and}
  \bibinfo{author}{\bibfnamefont{M.}~\bibnamefont{Dijkstra}},
  \bibinfo{journal}{J. Chem. Phys.} \textbf{\bibinfo{volume}{133}},
  \bibinfo{pages}{244115} (\bibinfo{year}{2010}).

\bibitem[{\citenamefont{van Megen and Underwood}(1993)}]{vanmegan1993}
\bibinfo{author}{\bibfnamefont{W.}~\bibnamefont{van Megen}} \bibnamefont{and}
  \bibinfo{author}{\bibfnamefont{S.~M.} \bibnamefont{Underwood}},
  \bibinfo{journal}{Nature (London)} \textbf{\bibinfo{volume}{362}},
  \bibinfo{pages}{616} (\bibinfo{year}{1993}).

\bibitem[{\citenamefont{Valeriani et~al.}(2011)\citenamefont{Valeriani, Sanz,
  Zaccarelli, Poon, Cates, and Pusey}}]{chantal2011}
\bibinfo{author}{\bibfnamefont{C.}~\bibnamefont{Valeriani}},
  \bibinfo{author}{\bibfnamefont{E.}~\bibnamefont{Sanz}},
  \bibinfo{author}{\bibfnamefont{E.}~\bibnamefont{Zaccarelli}},
  \bibinfo{author}{\bibfnamefont{W.}~\bibnamefont{Poon}},
  \bibinfo{author}{\bibfnamefont{M.}~\bibnamefont{Cates}}, \bibnamefont{and}
  \bibinfo{author}{\bibfnamefont{P.}~\bibnamefont{Pusey}}, \bibinfo{journal}{J.
  Phys.: Condens. Matter} \textbf{\bibinfo{volume}{23}},
  \bibinfo{pages}{194117} (\bibinfo{year}{2011}).

\bibitem[{\citenamefont{Bialk\'e et~al.}(2012)\citenamefont{Bialk\'e, Speck,
  and L\"owen}}]{lowen2012}
\bibinfo{author}{\bibfnamefont{J.}~\bibnamefont{Bialk\'e}},
  \bibinfo{author}{\bibfnamefont{T.}~\bibnamefont{Speck}}, \bibnamefont{and}
  \bibinfo{author}{\bibfnamefont{H.}~\bibnamefont{L\"owen}},
  \bibinfo{journal}{Phys. Rev. Lett.} \textbf{\bibinfo{volume}{108}},
  \bibinfo{pages}{168301} (\bibinfo{year}{2012}).

\bibitem[{\citenamefont{Ni}(2012)}]{nithesis}
\bibinfo{author}{\bibfnamefont{R.}~\bibnamefont{Ni}}, Ph.D. thesis,
  \bibinfo{school}{Utrecht University} (\bibinfo{year}{2012}).

\bibitem[{\citenamefont{Jepson et~al.}(2013)\citenamefont{Jepson, Martinez,
  Schwarz-Linek, Morozov, and Poon}}]{Jepson2013}
\bibinfo{author}{\bibfnamefont{A.}~\bibnamefont{Jepson}},
  \bibinfo{author}{\bibfnamefont{V.~A.} \bibnamefont{Martinez}},
  \bibinfo{author}{\bibfnamefont{J.}~\bibnamefont{Schwarz-Linek}},
  \bibinfo{author}{\bibfnamefont{A.}~\bibnamefont{Morozov}}, \bibnamefont{and}
  \bibinfo{author}{\bibfnamefont{W.~C.~K.} \bibnamefont{Poon}},
  \bibinfo{journal}{Phys. Rev. E} \textbf{\bibinfo{volume}{88}},
  \bibinfo{pages}{041002} (\bibinfo{year}{2013}).

\bibitem[{\citenamefont{Hall}(1972)}]{halleos}
\bibinfo{author}{\bibfnamefont{K.}~\bibnamefont{Hall}}, \bibinfo{journal}{J.
  Chem. Phys.} \textbf{\bibinfo{volume}{57}}, \bibinfo{pages}{2252}
  (\bibinfo{year}{1972}).

\bibitem[{\citenamefont{Ni et~al.}(submitted)\citenamefont{Ni, Cohen-Stuart,
  and Bolhuis}}]{ni2014}
\bibinfo{author}{\bibfnamefont{R.}~\bibnamefont{Ni}},
  \bibinfo{author}{\bibfnamefont{M.}~\bibnamefont{Cohen-Stuart}},
  \bibnamefont{and} \bibinfo{author}{\bibfnamefont{P.}~\bibnamefont{Bolhuis}},
  \bibinfo{journal}{arXiv:1403.1533}  (\bibinfo{year}{submitted}).

\bibitem[{\citenamefont{Schwarz-Linek et~al.}(2012)\citenamefont{Schwarz-Linek,
  Valeriani, Cacciuto, Cates, Marenduzzo, Morozov, and
  Poon}}]{schwarz-linek2012}
\bibinfo{author}{\bibfnamefont{J.}~\bibnamefont{Schwarz-Linek}},
  \bibinfo{author}{\bibfnamefont{C.}~\bibnamefont{Valeriani}},
  \bibinfo{author}{\bibfnamefont{A.}~\bibnamefont{Cacciuto}},
  \bibinfo{author}{\bibfnamefont{M.~E.} \bibnamefont{Cates}},
  \bibinfo{author}{\bibfnamefont{D.}~\bibnamefont{Marenduzzo}},
  \bibinfo{author}{\bibfnamefont{A.~N.} \bibnamefont{Morozov}},
  \bibnamefont{and} \bibinfo{author}{\bibfnamefont{W.~C.~K.}
  \bibnamefont{Poon}}, \bibinfo{journal}{Proc. Natl Acad. Sci. USA}
  \textbf{\bibinfo{volume}{109}}, \bibinfo{pages}{4052} (\bibinfo{year}{2012}).

\bibitem[{\citenamefont{van~der Meer et~al.}(2013)\citenamefont{van~der Meer,
  Fokkink, van~der Gucht, and Sprakel}}]{sprakel2013}
\bibinfo{author}{\bibfnamefont{B.}~\bibnamefont{van~der Meer}},
  \bibinfo{author}{\bibfnamefont{R.}~\bibnamefont{Fokkink}},
  \bibinfo{author}{\bibfnamefont{J.}~\bibnamefont{van~der Gucht}},
  \bibnamefont{and} \bibinfo{author}{\bibfnamefont{J.}~\bibnamefont{Sprakel}},
  \bibinfo{journal}{arXiv:1309.5837}  (\bibinfo{year}{2013}).

\bibitem[{\citenamefont{Hynninen et~al.}(2007)\citenamefont{Hynninen, Thijssen,
  Vermolen, Dijkstra, and Van~Blaaderen}}]{nanooptics}
\bibinfo{author}{\bibfnamefont{A.-P.} \bibnamefont{Hynninen}},
  \bibinfo{author}{\bibfnamefont{J.~H.~J.} \bibnamefont{Thijssen}},
  \bibinfo{author}{\bibfnamefont{E.~C.~M.} \bibnamefont{Vermolen}},
  \bibinfo{author}{\bibfnamefont{M.}~\bibnamefont{Dijkstra}}, \bibnamefont{and}
  \bibinfo{author}{\bibfnamefont{A.}~\bibnamefont{Van~Blaaderen}},
  \bibinfo{journal}{Nature Materials} \textbf{\bibinfo{volume}{6}},
  \bibinfo{pages}{202} (\bibinfo{year}{2007}).

\bibitem[{\citenamefont{Eldridge et~al.}(1993)\citenamefont{Eldridge, Madden,
  and Frenkel}}]{ab13}
\bibinfo{author}{\bibfnamefont{M.}~\bibnamefont{Eldridge}},
  \bibinfo{author}{\bibfnamefont{P.}~\bibnamefont{Madden}}, \bibnamefont{and}
  \bibinfo{author}{\bibfnamefont{D.}~\bibnamefont{Frenkel}},
  \bibinfo{journal}{Nature (London)} \textbf{\bibinfo{volume}{365}},
  \bibinfo{pages}{35} (\bibinfo{year}{1993}).

\end{thebibliography}

\end{document}